\documentclass[prl,twocolumn,showpacs,preprintnumbers,amsmath,amssymb,superscriptaddress]{revtex4}
\usepackage{graphicx}% Include figure files
\usepackage{dcolumn}%  Align table columns on decimal point
\usepackage{bm}%       bold math

\usepackage{color}

\allowdisplaybreaks[1]

\begin{document}
%%%%%%%%%%%%%%%%

%\preprint{APS/123-QED}

%%%%%%%%%%%%%%%%%%%%%%%%%%%%%%%%%%%%%%%%%%%%%%%%%%%%%%%%%%%%%%%%%%%%%%%%
\title{
First-principles calculation of the instability leading to giant inverse
magnetocaloric effects
}
%%%%%%%%%%%%%%%%%%%%%%%%%%%%%%%%%%%%%%%%%%%%%%%%%%%%%%%%%%%%%%%%%%%%%%%%

\author{D.\ Comtesse}
\affiliation{Faculty of Physics and Center for Nanointegration, CENIDE,
             University of Duisburg-Essen, D-47048 Duisburg, Germany}  
\author{M.\ E.\ Gruner}
\affiliation{Faculty of Physics and Center for Nanointegration, CENIDE,
             University of Duisburg-Essen, D-47048 Duisburg, Germany}  
\author{M.\ Ogura}
\affiliation{Department of Physics, Graduate School of Science, 
             Osaka University, Machikaneyama 1-1, Toyonaka, 
             Osaka 560-0043, Japan}
\author{V.\ V.\ Sokolovskiy}
\affiliation{National University of Science and Technology, 'MIS\&S',
             119049 Moscow, Russia}
\affiliation{Condensed Matter Physics Department, 
             Chelyabinsk State University, 454021 Chelyabinsk, Russia}
\author{V.\ D.\ Buchelnikov}
\affiliation{Condensed Matter Physics Department, 
             Chelyabinsk State University, 454021 Chelyabinsk, Russia}
\author{A.\ Gr\"unebohm}
\affiliation{Faculty of Physics and Center for Nanointegration, CENIDE,
             University of Duisburg-Essen, D-47048 Duisburg, Germany}  
\author{R.\ Arr\'oyave}
\affiliation{Department of Materials Science and Engineering, Texas A\&M University,
           College Station, Texas 77843, USA}
\author{N.\ Singh}
\affiliation{Department of Engineering Technology, University of Houston,
           Houston, Texas 77204, USA}
\author{T.\ Gottschall}
\affiliation{Materials Science, Technical University Darmstadt, D-64287 Darmstadt, Germany}
\author{O.\ Gutfleisch}
\affiliation{Materials Science, Technical University Darmstadt, D-64287 Darmstadt, Germany}
\author{V.\ A.\ Chernenko}
\affiliation{BCMaterials,University of Basque Country (UPV/EHU)and Ikerbasque, 
           Basque Foundation for Science, Bilbao 48011, Spain}
%Departamento de Electricidad y Electronica,
%             Universidad del Pais Vasco, UPV/EHU, 48080 Bilbao, Spain}
\author{F.\ Albertini}
\affiliation{IMEM-CNR, Parco Area delle Scienze 37/A, I-43124 Parma, 
             Italy}
\author{S.\ F\"ahler}
\affiliation{IFW Dresden, P.\ O.\ Box 270116, D-01171 Dresden, Germany}
\author{P.\  Entel}\email{entel@thp.uni-duisburg.de}
\affiliation{Faculty of Physics and Center for Nanointegration, CENIDE,
             University of Duisburg-Essen, D-47048 Duisburg, Germany}

\date{\today}

%%%%%%%%%%%%%%%%
\begin{abstract}
%%%%%%%%%%%%%%%%

The structural and magnetic properties of functional 
Ni-Mn-Z (Z = Ga, In, Sn) Heusler alloys are studied by first-principles 
and Monte Carlo methods. The \textit{ab initio} calculations give a basic 
understanding of the underlying physics which is associated with the 
strong competition of ferro- and antiferromagnetic interactions with increasing 
chemical disorder. The resulting $d$-electron orbital dependent magnetic ordering 
is the driving mechanism of magnetostructural instability which is accompanied 
by a drop of magnetization governing the size of the magnetocaloric 
effect. The thermodynamic properties are calculated by using the 
\textit{ab initio} magnetic exchange coupling constants in finite-temperature 
Monte Carlo simulations, which are used to accurately reproduce the experimental 
entropy and adiabatic temperature changes across the magnetostructural transition. 

%%%%%%%%%%%%%%
\end{abstract}
%%%%%%%%%%%%%%


\maketitle

Following the concepts of Hume-Rothery the influence of composition on
martensitic and magnetic transformation temperatures is commonly condensed 
as a dependency of electrons per atom ($e/a$-ratio) \cite{chernenko-1999}.
Experiment and first-principles calculations, however, reveal that the
Z element in Ni-Mn-Z Heusler alloys (Z = Ga, In, Sn) also affects the transformation 
temperatures substantially \cite{entel-epjb-2013}. Moreover, recent experiments on 
samples with identical composition but different heat treatment indicate that 
chemical disorder also plays an important role \cite{ito-2007,kustov-2009,niemann-2012a}. 
Here, we use first-principles calculations to identify the influence of chemical
disorder on the magnetic exchange parameters and derive guidelines for a further 
systematic improvement of magnetocaloric materials \cite{sandeman-2012}.

Besides the magnetocaloric effect (MCE) in Gd and other alloys  at room
temperature \cite{gschneidner-1997,tegus-2002}, the metamagnetic Ni-Mn based Heusler 
materials \cite{krenke-2005,kainuma-2006}, have attracted much interest recently 
\cite{planes-2009,acet-2011}. In these alloys the metamagnetic features are responsible 
for magnetic glass behavior and frustration due to chemical disorder 
\cite{leighton-2012,cong-2012,chaddah-2012} as well as unusual magnetization behavior 
under an external magnetic field such as a large jump of the magnetization 
$\Delta M(T_m)$ at the martensitic/magnetostructural transformation temperature 
$T_m$ \cite{barandiaran-2013}. This gives rise to the large inverse MCE of the
materials \cite{krenke-2005,kainuma-2006,buche-2008,buche-2011}. 
The MCE can be influenced when Ni is substituted in part by Co: It is strongly enhanced 
in the case of In-based intermetallics \cite{bourgault-2010a,gutfleisch-2012}
(with adiabatic temperature change $\Delta T_{ad} = -6$ K in 2 T field 
\cite{gutfleisch-2012}) while in the case of Ga the MCE is turned from direct to inverse
by decoupling $T_m$ and Curie temperature $T_C$ \cite{fabbrici-2011} (with 
$\Delta T_{ad} = -1.6$ K in 1.9 T field \cite{franca-2012a,franca-2012b}).

Chemical disorder in the Mn-rich Heusler alloys is responsible for competing magnetic 
interactions (ferromagnetic versus antiferromagnetic) because the extra Mn atoms occupy 
lattice sites of the Z-sublattice which interact antiferromagnetically with the Mn atoms  
on the Y-sublattice due to RKKY-type interactions. This competition of magnetic
interactions leads to the characteristic drop of magnetization curves at $T_m$,
which is observed in Ni-(Co)-Mn-Z materials
\cite{cong-2012,chaddah-2012,barandiaran-2013,bourgault-2010a,gutfleisch-2012,fabbrici-2011,franca-2012a,franca-2012b}.
(The magnetostructural transformation was originally discussed for Ni-excess 
Ni-Mn-Ga alloys which was interpreted as magnetic dilution effect \cite{khovaylo-2005}.)

%%%%%%%%%%%%%%%%%%%
\begin{figure}[htb]
%%%%%%%%%%%%%%%%%%%
%
\centerline{
  \includegraphics[width=4.25cm,clip]{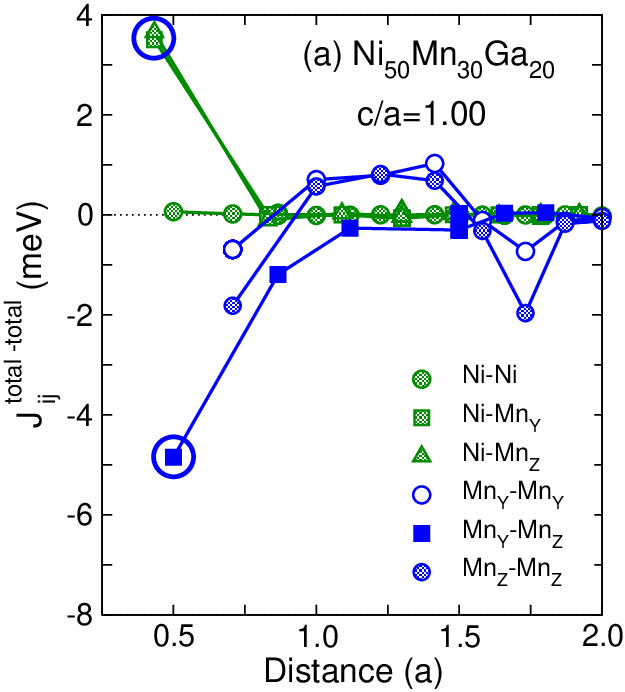}
  \hspace{0.0cm}
  \includegraphics[width=4.25cm,clip]{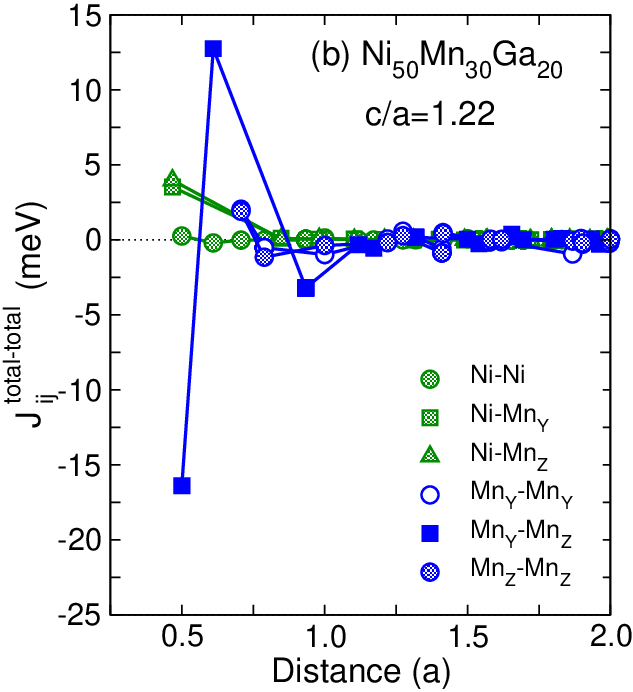}
           }
\vspace*{0.1cm}
\centerline{
  \includegraphics[width=4.25cm,clip]{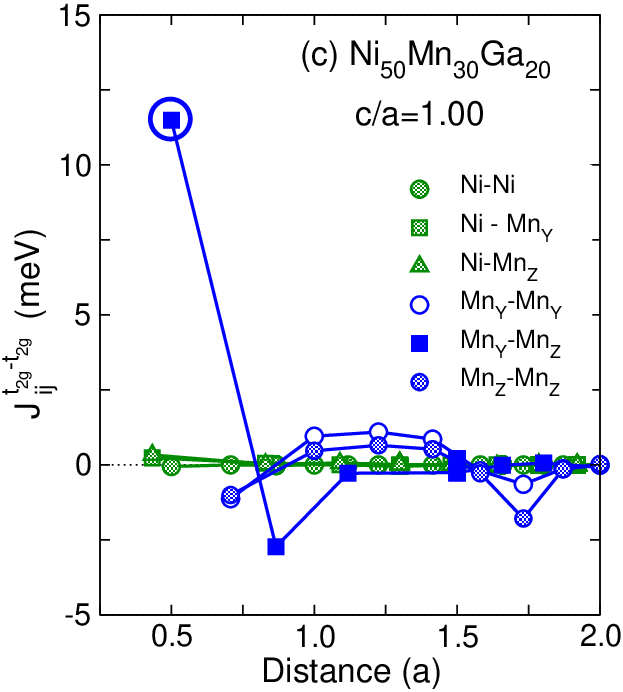}
  \hspace{0.0cm}
  \includegraphics[width=4.25cm,clip]{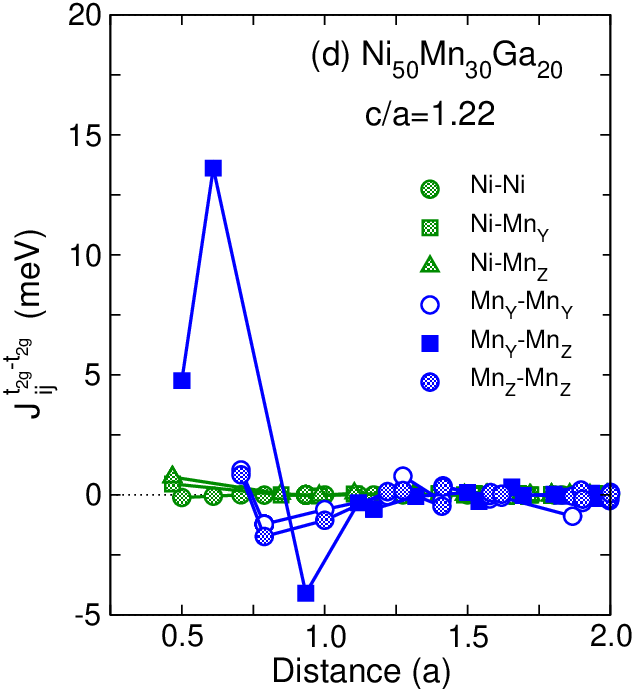}
           }
\vspace*{0.0cm}
\centerline{
  \includegraphics[width=4.25cm,clip]{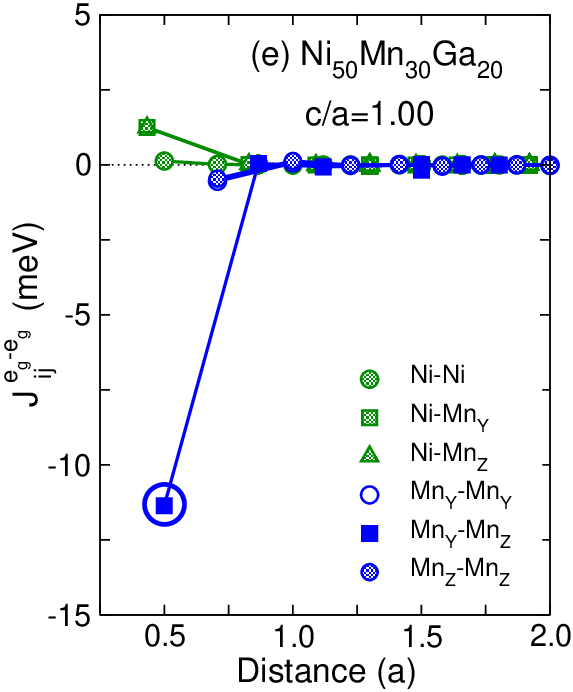}
  \hspace{0.1cm}
  \includegraphics[width=4.25cm,clip]{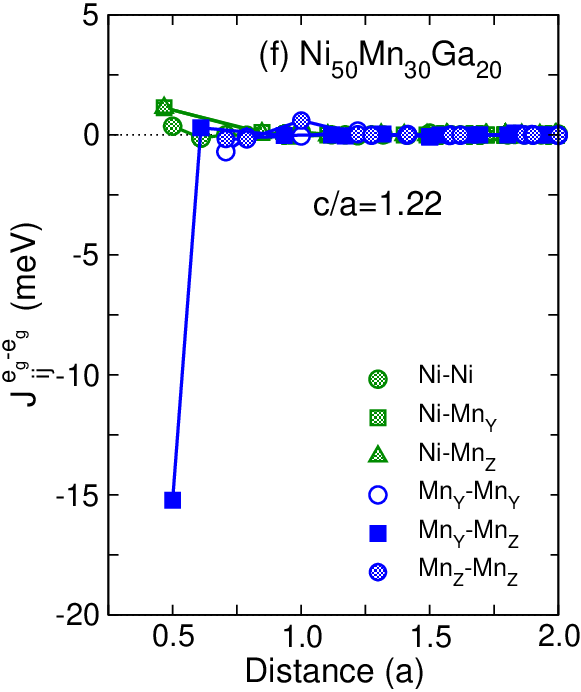}
           }
\caption{
         (a-f) Element and orbital resolved magnetic exchange parameters of austenite and 
         martensite Ni$_{50}$Mn$_{30}$Ga$_{20}$ from \textit{ab initio} 
         calculations (not all contributions are shown). It is obvious that the 
         effect of ferromagnetic interactions is largely compensated by the influence 
         of the antiferromagnetic interactions. (Austenite: $a = 5.85$ {\AA}, martensite: 
         $a = b = 5.47$ {\AA}, $c = 6.68$, {\AA}, $c/a = 1.22$.)
         }
\label{figure-1}
%
%%%%%%%%%%%%
\end{figure} 
%%%%%%%%%%%%

Previously, Monte Carlo (MC) methods  \cite{buche-2008,buche-2011} and first-principles 
tools \cite{vasp,spr-kkr,masako-code} have been used to explore the phase
diagram and magnetostructural transformation of the magnetocaloric materials 
as a function of temperature and $e/a$ \cite{entel-epjb-2013,uijttewaal-2009}.

%%%%%%%%%%%%%%%%%%%
\begin{figure}[htb]
%%%%%%%%%%%%%%%%%%%
\centerline{
  \includegraphics[width=8.5cm,clip]
      {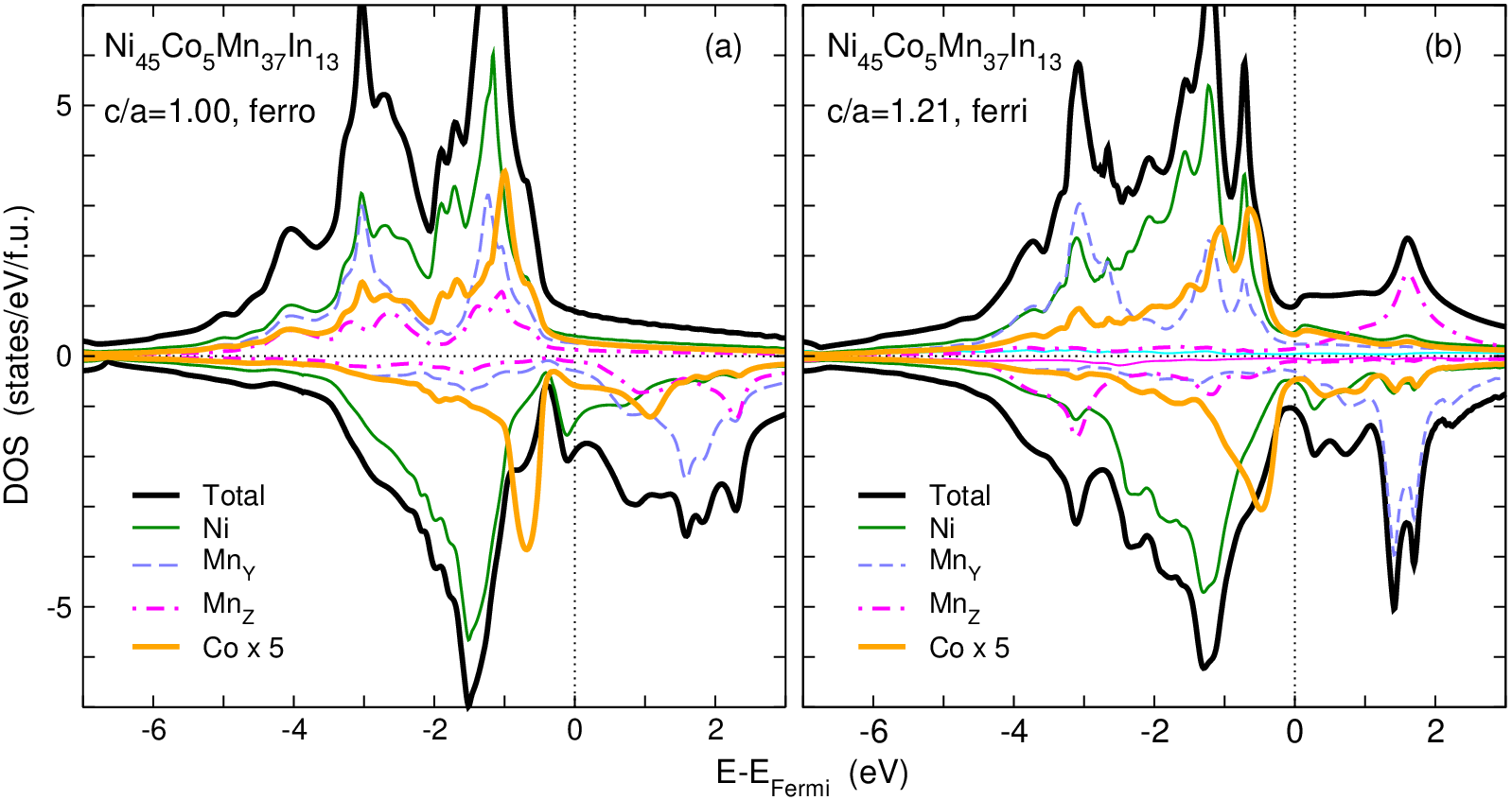}
           }
\caption{
         Element resolved density of states of (a) austenite
         (L2$_1$, $a = 5.96$ {\AA}) and (b) martensite 
         (L1$_0$, $a =b =5.6$ {\AA}, $c = 6.76$ {\AA}, $c/a = 1.21$) of
         Ni$_{45}$Co$_5$Mn$_{37}$In$_{13}$ showing the stabilization of martensite 
         due to the formation of a pseudogap at $E_F$ (``ferri'' means that the spin of 
         Mn on the In sites is reversed). 
         }
\label{figure-2}
%
%%%%%%%%%%%%
\end{figure} 
%%%%%%%%%%%%

In this letter we show that basically the magnetic exchange interactions obtained
from first-principles calculations allows to calculate the MCE as a function of 
temperature across the magnetostructural transition. 

Before addressing the MCE, we would like to highlight the complex magnetic behavior 
of the disordered Ni-(Co)-Mn-Z alloys with excess Mn. We evaluate the  effective exchange 
coupling constants $J_{ij}$ using the KKR CPA method \cite{spr-kkr,masako-code} where, 
following the prescription of \cite{lichtenstein-1987}, the $J_{ij}$ are obtained from
%
%%%%%%%%%%%%%%
{\begin{align}
%%%%%%%%%%%%%%
%
J_{ij} & =  \frac{1}{4\pi}\int_{-\infty}^{\epsilon_F} dE\, {\mathrm{Im}} \,
            {\mathrm{Tr}} \left[ \Delta_i 
            \tau_{\uparrow}^{ij} \Delta_j \tau_{\downarrow}^{ji} \right] ,
%
%%%%%%%%%%%
\end{align}
%%%%%%%%%%%
%
where $\Delta_i$ is the difference in the inverse single-site scattering $t$-matrices 
for spin-up and spin-down states, 
$\Delta_i = t_{i\uparrow}^{-1} - t_{i\downarrow}^{-1}$, and $\tau$ is the
scattering path operator. Since we use the spherical potential and scalar
relativistic approximation, the $t$-matrices are diagonal. Thus, we can decompose 
the $J_{ij}$ and extract the contribution between $L$ states ($L = (l,m)$ indicates the
set of angular momentum and magnetic quantum number) at the $i$-th site and $L^{\prime}$ 
states at the $j$-site as
%
%%%%%%%%%%%%%
\begin{align}
%%%%%%%%%%%%%
%
J_{ij}^{L-L^{\prime}} & = \frac{1}{4\pi} 
           \int_{-\infty}^{\epsilon_F} dE \, {\mathrm{Im}}
           \Delta_{iL} \tau_{\uparrow LL^{\prime}}^{ij}
           \Delta_{jL^{\prime}} \tau_{\downarrow L^{\prime}L}^{ji} \, .
%
%%%%%%%%%%%
\end{align}
%%%%%%%%%%%

This allows to calculate element as well as orbital resolved magnetic 
coupling constants as shown, e.g., in Fig.\ \ref{figure-1} for disordered, non-stoichiometric 
Ni$_{50}$Mn$_{30}$Ga$_{20}$. (The effect of disorder on martensitic 
transformation has also been discussed by \cite{kulkova-2011}.)

%%%%%%%%%%%%%%%%%%%%
\begin{figure*}[htb]
%%%%%%%%%%%%%%%%%%%%
\centerline{
  \includegraphics[width=8.0cm,clip]{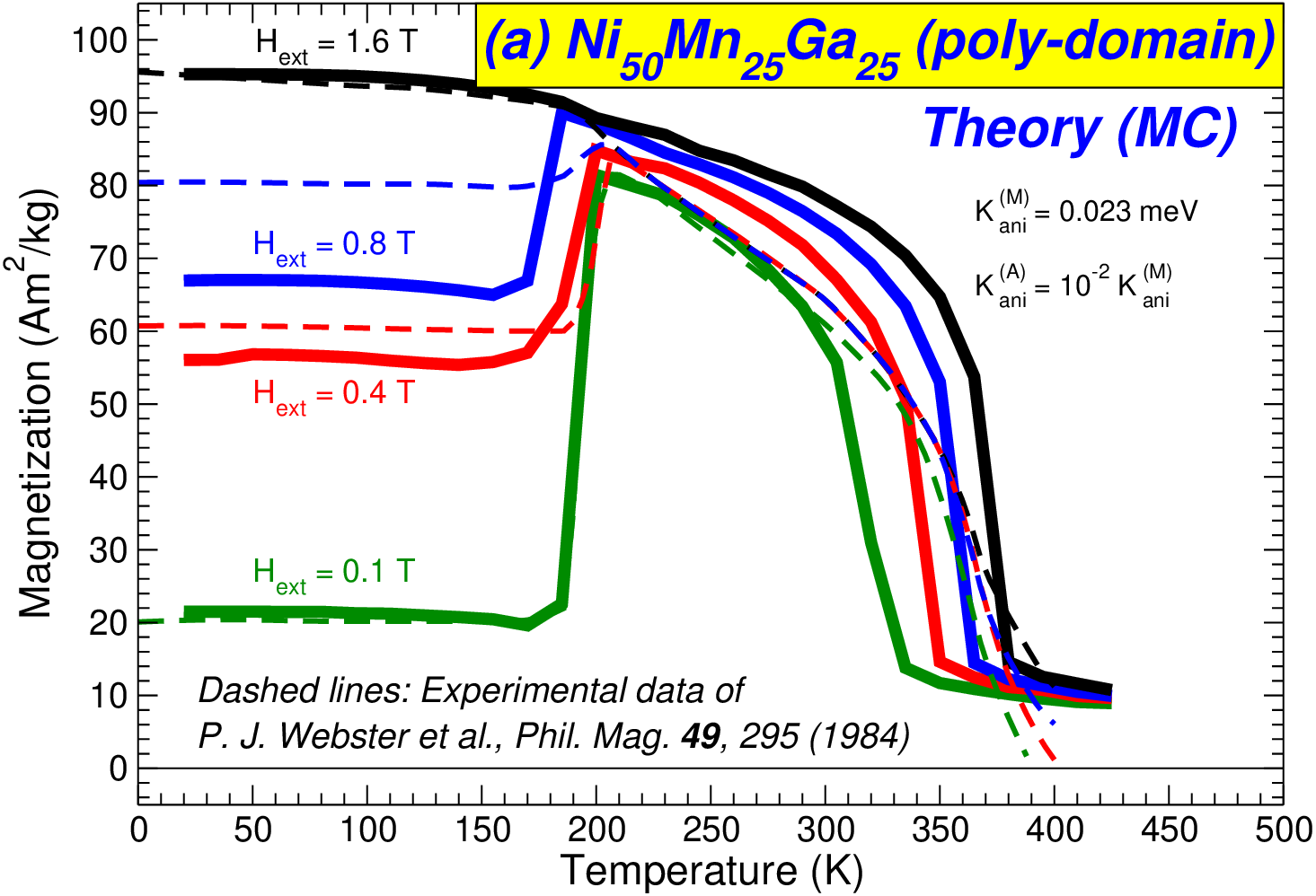}
  \hspace{0.3cm}
  \includegraphics[width=8.0cm,clip]{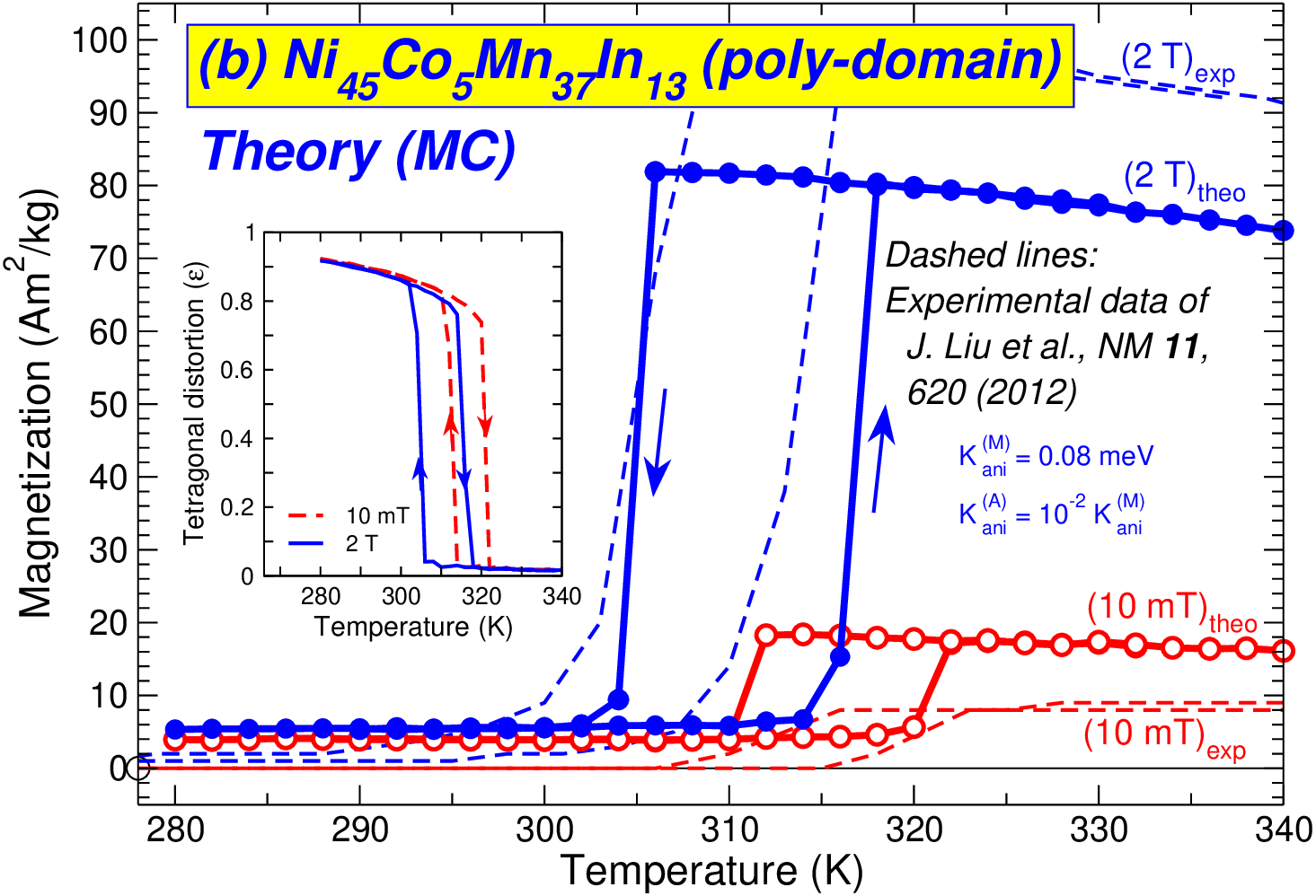}
           }
\caption{
         (a) Monte Carlo magnetization curves (solid lines) of 
         Ni$_{50}$Mn$_{25}$Ga$_{25}$ from the Potts model using \textit{ab initio} 
         magnetic exchange and magnetocrystalline anisotropy parameters, in comparison 
         with experiment \cite{webster-1984}. The jump of the magnetization curves 
         vanishes for sufficiently large fields.
         (b) Theoretical and experimental magnetization curves 
         of Ni$_{45}$Co$_5$Mn$_{37}$In$_{13}$ in a small and large external magnetic 
         field of 10 mT and 2 T \cite{gutfleisch-2012}. Here, the jump $\Delta M$ does 
         not vanish for increasing magnetic fields. The inset shows the hysteresis 
         from the first-order cubic-tetragonal transformation.
         }
\label{figure-3}
%
%%%%%%%%%%%%%
\end{figure*} 
%%%%%%%%%%%%%

From the behavior of the magnetic coupling constants in Fig.\ \ref{figure-1} 
for austenite and martensite Ni$_{50}$Mn$_{30}$Ga$_{20}$ as a function of the 
distance between the atoms we notice the nearly perfect compensation of ferromagnetic 
interactions associated with the more itinerant $d$-electron $t_{2_g}$ states and 
antiferromagnetic interactions of magnetic moments associated with the more localized 
$e_{g}$ states. This destabilizes ferromagnetic austenite which undergoes 
a magnetostructural transformation to paramagnetic martensite
(after addition of Co \cite{fabbrici-2011}). The magnetic exchange coupling 
constants of tetragonally distorted martensite ($c/a = 1.22$) 
cannot stabilize ferromagnetic order any more. In the MC simulations we use the
magnetic exchange parameters from the zero-temperature \textit{ab initio} calculations 
for austenite and martensite which we let merge at $T_m$. Thermal spin fluctuations will 
further help to stabilize the ``paramagnetic'' gap at $T < T_m$ below the magnetic 
jump $\Delta M(T_m)$ \cite{franca-2012a,franca-2012b}. 
This shows that the martensitic/magnetostructural instability is to a large extent driven 
by the atomic disorder leading to strong competition of 
ferro- and antiferromagnetic interactions of approximately equal strength. 
This is responsible for the breakdown of ferromagnetic order in austenite and the appearance 
of the ``paramagnetic gap'' below $T_m$ in martensite. 

We emphasize that the same scenario works in case of Ni-(Co)-Mn-(In, Sn). With addition 
of Co (Co is likely to replace Ni because of similar coordination chemistry), the 
ferromagnetic tendencies increase but also the ``disordered nature'' of the magnetic 
interactions, which governs the magnetostructural instability. 
Although, Co leads to a decrease of $T_m$, it clearly has a favorable effect regarding 
the magnetocaloric properties. This is because Co hybridizes strongly with Ni states 
(see Fig.\ \ref{figure-2}) and causes more spin disorder in the Heusler materials 
which leads to larger magnetic entropy and adiabatic temperature changes. (Note that 
competing interactions have also been discussed for Ni-Mn-Ga in 
\cite{galanakis-2011,patricia-njp}.)

The jump $\Delta M(T_m)$ which is accompanied by breakdown of long-range
ferromagnetism, large magnetic fluctuations and a large entropy change across the 
magnetostructural transition, is at the heart of the giant inverse MCE. 

Regarding thermodynamic properties we use the Blume-Emery-Griffiths (BEG) model 
\cite{castan-1999} for austenite-martensite transformation in combination with the 
Potts model for the magnetic part and a magnetoelastic interaction term 
\cite{buche-2008,buche-2011}:  
${\cal H} = {\cal H}_m + {\cal H}_{el} + {\cal H}_{int}$ where
%
%%%%%%%%%%%%%%%%
\begin{eqnarray}
%%%%%%%%%%%%%%%%
%
%{\cal H} & = &  {\cal H}_m + {\cal H}_{el} + {\cal H}_{int}, 
%
{\cal H}_m & = & {}
       -  \sum_{\langle ij \rangle} J_{i,j} \, \delta_{S_i,S_j} 
       - g \mu_B H_{ext} \sum_i \, \delta_{S_i, S_g} M_i
 \nonumber \\ && {}
       + K_{ani} \sum_i \, \delta_{S_i, S_k} M_i^2 \, ,
\\*[0.1cm]
{\cal H}_{el} & = & {} - \sum_{\langle ij \rangle} \sigma_i \sigma_j
           \left( 
           J + U_{1}g \mu_B H_{ext} \sum_i \delta_{\sigma_i \sigma_g} 
           \right) 
\nonumber \\ && {}
         - K \sum_{\langle ij \rangle} 
           \left( 1-\sigma_i^2 \right) \left( 1-\sigma_j^2 \right) .
%
%%%%%%%%%%%%%%
\end{eqnarray}
%%%%%%%%%%%%%%

The $J_{i,j}$ are the magnetic exchange parameters for each structure. Since mapping 
of \textit{ab initio} energies is only onto $J_{ij}$ with unit length of spins, field 
terms must include explicitely $M_i, \; M_i^2$, where $M_i$ is the \textit{ab initio} 
value of magnetization of atom at site $i$ taken to be dimensionless. $J$ and $K$ are 
the elastic and $U_{1}$ ($U_{i,j}$) the magnetoelastic interaction parameters 
(${\cal H}_{int}$ couples $M_iM_j$ and $\sigma_i^2$, $\sigma_j^2$ with strength $U_{ij}$ 
\cite{buche-2008,castan-1999}). The Kronecker symbol restricts the spin-spin interactions 
to those between the same Potts-$q$ states. %%% defined at lattice site $i = 1 \ldots N$.
The spin moment of Mn is $S = \textstyle\frac{5}{2}$ and we identify the $2S + 1$ spin 
projections with $q_{Mn} = 1 \ldots 6$. Likewise, we assume $S=1$ for Ni and 
$S = \textstyle\frac{3}{2}$ for Co. The BEG model defines $\sigma_i = 0, \pm 1$ for 
austenite and two martensitic variants, respectively 
\cite{buche-2008,buche-2011,castan-1999}; 
it allows first-order martensitic phase transformation with thermal hysteresis for 
sufficiently large biquatratic elastic interaction  ($0.2 <  K/J < 0.37$; we have adopted
0.23 for Ni-(Co)-Mn-In alloys).
Because of the magnetoelastic coupling term the jump $\Delta M(T_m)$ is coupled to 
the martensitic transformation and exhibits hysteresis as well. 

We mimic the magnetic aspect of polycrystalline materials by magnetic domain
blocks with random initial spin configurations in each domain. The spins from different 
domain blocks can interact with probability 
$W = \textnormal{min}(1,\textnormal{exp}(-|K_{ani}| M_i^2/g \mu_B H_{ext} |M_i|)$.
This stochastic competition between the magnetic anisotropy field and external magnetic 
field allows to realize experimental trends of magnetization curves, see Fig.\ 
\ref{figure-3}. 

Hence, the extended Potts model in  Eqs.\ (3-4) allows us to describe magnetic, 
structural as well as coupled magnetostructural phase transitions  
\cite{buche-2008,buche-2011}. Figure \ref{figure-3} shows magnetization curves in small 
and large magnetic fields for polycrystalline Ni$_{50}$Mn$_{25}$Ga$_{25}$ and
Ni$_{45}$Co$_5$Mn$_{37}$In$_{13}$ in comparison to experiment 
\cite{webster-1984,gutfleisch-2012}.

Note that there is a distinctive difference between $M(T)$ of the two materials.

For ferromagnetic Ni-Mn-Ga alloys near stoichiometry the jump vanishes for 
sufficiently large magnetic field when overcoming the 
magnetocrystalline anisotropy in agreement with experiment 
\cite{planes-2009,acet-2011,webster-1984}. However, the jump 
persists for the Mn-rich Ni-Co-Mn-(Ga, In, Sn) up to large magnetic fields
\cite{bourgault-2010a,gutfleisch-2012,fabbrici-2011,franca-2012a,franca-2012b} 
because of strong antiferromagnetic Mn$_Y$-Mn$_Z$ interactions competing with the 
ferromagnetic ones. Note that for Ni$_{45}$Co$_5$Mn$_{37}$In$_{13}$ magnetization jump 
and hysteretic behavior in Fig.\ \ref{figure-3}(b) agree well with experiment 
\cite{gutfleisch-2012}.

Total enery calculations and MC simulations show that the magnetostructural 
instability observed in Mn-rich systems 
\cite{cong-2012,chaddah-2012,barandiaran-2013,bourgault-2010a,gutfleisch-2012,fabbrici-2011,franca-2012a,franca-2012b}
is accompanied by a transition from ferromagnetic austenite to ferrimagnetic/paramagnetic
martensite, although, zero-temperature energy differences, $(E_{ferro}-E_{ferri})$ may 
already become small in austenite (this is the case for Ni$_{50-x}$Co$_x$Mn$_{25+y}$Sn$_{25-y}$ 
at a critical Co concentration). Free energies are difficult to evaluate  on an 
\textit{ab initio} basis because of chemical disorder, softenig of lattice vibrations
in austenite and magnetic excitations, which is beyond the scope of the present paper.

The adiabatic temperature changes across Curie temperature (direct MCE) and across 
magnetostructural transformation (inverse MCE) are determined by the isothermal magnetic 
entropy change and total specific heat (sum of magnetic and lattice specific heat, where 
the latter part is taken from the Debye model). These quantities can be calculated from the
relations:
%
%%%%%%%%%%%%%
\begin{align}
%%%%%%%%%%%%%
%
\Delta T_{ad}(T, H_{ext}) & = 
                  -\mu_0 \! \int\limits_{0}^{H_{ext}} \! dH^{\prime} \,
                  \frac{T}{C(T,H^{\prime})} 
                  \left(
                  \frac{\partial M}{\partial T}
                  \right)_{\! H^{\prime}}
\nonumber \\ {} & \approx 
                  -T \frac{\Delta S_{mag}(T, H_{ext})}{C(T, H_{ext})} \, ,
\\ 
\Delta S_{mag}(T_m, H_{ext}) & =  \Delta M(T_m, H_{ext}) 
               \left(\frac{dT_m}{dH_{ext}}\right)^{\! -1} \! \! \! ,
\\
\Delta T_{ad}(T, H_{ext}) =  -T & \frac{\Delta S_{mag}(T_m, H_{ext})}{C(T,H_{ext})}
                      \Delta f(T, H_{ext}) 
%
%%%%%%%%%%%
\end{align}
%%%%%%%%%%%
%
Here, $\Delta S_{mag}(T, H_{ext}) =  S_{mag}(T, H_{ext}) - S_{mag}(T, 0)$
is the entropy difference for finite and zero field.  We use the Maxwell
relation in Eq.\ (5) for the direct MCE while for the inverse MCE at the first-order 
magnetostructural transition we use instead Eqs.\ (6) and (7) based on the 
Clausius-Clapeyron equation. $\Delta M$ and $\Delta S_{mag}$ are the jump of 
magnetization and magnetic entropy, $dT_m/dH_{ext}$ is the shift of structural phase 
transition in the magnetic field and $\Delta f(T,H_{ext})$ is the change of austenite 
fraction caused by a field change, and $C(T,H_{ext})$ is the total specific heat. In 
our calculations the value of $dT_m/dH_{ext}$ was taken from experiment 
\cite{gutfleisch-2012}, whereas $\Delta M$ and $\Delta f$ were obtained from MC 
simulations.

%%%%%%%%%%%%%%%%%%%
\begin{figure}[htb]
%%%%%%%%%%%%%%%%%%%
\centerline{
  \includegraphics[width=7.5cm,clip]{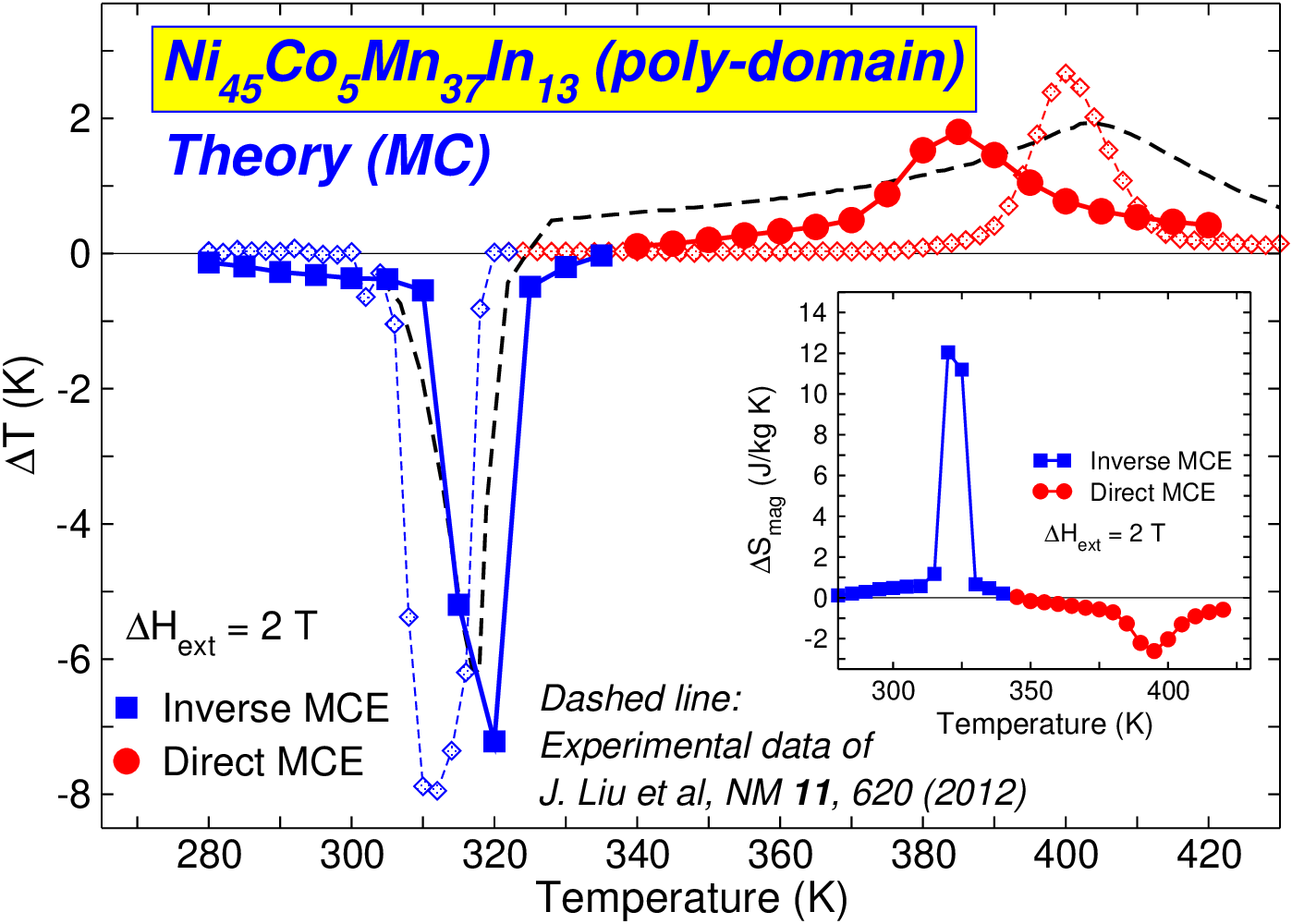}
           }
\caption{
         Adiabatic temperature change of the direct (red circles) and inverse 
         (blue squares) MCE of polycrystalline Ni$_{45}$Co$_5$Mn$_{37}$In$_{13}$ 
         as obtained from the extended Potts model (diamonds: results for single-domain state; 
         inset: temperature variation of the entropy). 
         }
\label{figure-4}
%
%%%%%%%%%%%%
\end{figure} 
%%%%%%%%%%%%

Results for the MCE of Ni-Co-Mn-In alloys are shown in Fig.\ \ref{figure-4} which 
demonstrates the enhancement of cooling up to -6 K compared to samples without Co 
($\approx -3$ K) \cite{gutfleisch-2012}. The enhancement is caused by the increased 
magnetic disorder leading to larger $\Delta S_{mag} (T, H_{ext})$. However, systematic 
comparison is difficult because of either different sample
preparation or compositions, compare, for instance, the different values reported
for $\Delta S_{mag}$ and $\Delta T_{ad}$ regarding Ni-Mn-In alloys 
\cite{bourgault-2010a,gutfleisch-2012,moya-2007}.
The key figure of merit in these alloys is large: we obtain 
$\mathrm{RCP}_{inv} = -132$ J/kg for Ni$_{45}$Co$_5$Mn$_{37}$In$_{13}$.

In this paper we have shown that MCE is determined by the influence 
of competing ferromagnetic and antiferromagnetic interactions in 
Ni-Co-Mn-Z  alloys \cite{entel-epjb-2013}. 
The \textit{ab initio} magnetic exchange parameters are the decisive parameters which 
determine the jump $\Delta M(T_m, H_{ext})$ and the size of the MCE. From the predictive 
power of \textit{ab initio} calculations regarding the influence of the magnetic coupling 
constants to optimize the MCE, it might be worth to create even more spin disorder and 
larger isothermal entropy changes by looking for the effects of Cr and Gd added to 
Ni-Co-Mn-Z materials. 

As already indicated by several experiments 
\cite{chernenko-1999,ito-2007,kustov-2009,niemann-2012a}
our calculations show that $e/a$ ratios are not sufficient for describing the 
transformation behavior (like disorder broadened first-order magnetostructural phase 
transition, range of coexistence of phases and metastability 
\cite{cong-2012,chaddah-2012,barandiaran-2013,bennett-2012} 
and magnetic cluster formation \cite{leighton-2012}) 
and MCE in Heusler alloys completely. Though 
this concept of itinerant electrons gives a rough overview on the transformation
temperatures, the interaction of the localized electronic orbitals influences the 
exchange parameters and thus the size of the magnetocaloric effect. In particular, 
our calculations identify that it is beneficial having specific chemical environment for 
the Mn$_Y$ and Mn$_Z$ atoms since this optimizes the compensation of ferromagnetic 
and antiferromagnetic interactions. As chemical order is susceptible to time and 
temperature during the sample preparation in addition to composition, this guideline 
will allow for a systematic optimization of magnetocaloric materials.

%%%%%%%%%%%%%%%%%%%%%%%
\begin{acknowledgments}
%%%%%%%%%%%%%%%%%%%%%%%
%
We thank the DFG (SPP 1599) for financial support. RA and NS acknowlege support from 
NSF through Grants DMR-0844082 and 0805293.
%
%%%%%%%%%%%%%%%%%%%%%
\end{acknowledgments}
%%%%%%%%%%%%%%%%%%%%%

%%%%%%%%%%%%%%%%%%%%%%%%%%%

%%%%%%%%%%%%%%%%%%%%%

%%%%%%%%%%%%%%
\end{document}